\documentclass[fleqn,twoside,twocolumn,nofootinbib,showkeys]{revtex4} 
\usepackage[sec,doi]{ujp_UTF8} 

\begin{document}
\title[MOCHI in UHECR showers]
{Modified Characteristics of Hadronic Interactions in Ultra-high-energy Cosmic-ray Showers}%
\author{J.~Ebr}
\affiliation{FZU -- Institute of Physics of the Czech Academy of Sciences}
\address{Na Slovance 1999/2, Prague, Czech Republic}
\email{ebr@fzu.cz}
\author{J. Blažek}
\affiliation{FZU -- Institute of Physics of the Czech Academy of Sciences}
\address{Na Slovance 1999/2, Prague, Czech Republic}
\author{J. Vícha}
\affiliation{FZU -- Institute of Physics of the Czech Academy of Sciences}
\address{Na Slovance 1999/2, Prague, Czech Republic}
\author{T. Pierog}
\affiliation{Karlsruhe Institute of Technology}
\address{76131 Karlsruhe, Germany}
\author{E. Santos}
\affiliation{FZU -- Institute of Physics of the Czech Academy of Sciences}
\address{Na Slovance 1999/2, Prague, Czech Republic}
\author{P. Trávníček}
\affiliation{FZU -- Institute of Physics of the Czech Academy of Sciences}
\address{Na Slovance 1999/2, Prague, Czech Republic}
\author{N. Denner}
\affiliation{FZU -- Institute of Physics of the Czech Academy of Sciences}
\address{Na Slovance 1999/2, Prague, Czech Republic}
\author{R. Ulrich}

\udk{№ УДК/UDC} \razd{\secix}

\autorcol{J.~Ebr, J.~Blažek, J.~Vícha et al.}%

\setcounter{page}{1}%

\begin{abstract}
Data from multiple experiments suggest that the current interaction models used in Monte Carlo simulations do not correctly reproduce the hadronic interactions in air showers produced by ultra-high-energy cosmic rays (UHECR), in particular – but not limited to – the production of muons during the showers. We have created a large library of UHECR simulations where the interactions at the highest energies are slightly modified in various ways – but always within the constraints of the accelerator data, without any abrupt changes with energy and without assuming any specific mechanism or dramatically new physics at the ultra-high energies. We find that even when very different properties – cross-section, elasticity and multiplicity – of the interactions are modified, the resulting changes in some air-shower observables are still mutually correlated. Thus not all possible combinations of changes of observables are easily reproduced by some combination of the modifications. Most prominently, the recent results of the Pierre Auger Observatory, which call for a change in the prediction of both the muon content at ground and the depth of the maximum of longitudinal development of the showers, are rather difficult to reproduce with such modifications, in particular when taking into account other cosmic-ray data. While some of these results are related to the assumptions we place on the modifications, the overall lessons are general and provide valuable insight into how the UHECR data can be interpreted from the point of view of hadronic physics. 
\end{abstract}


\maketitle

\begin{figure}
\vskip1mm
\includegraphics[width=\column]{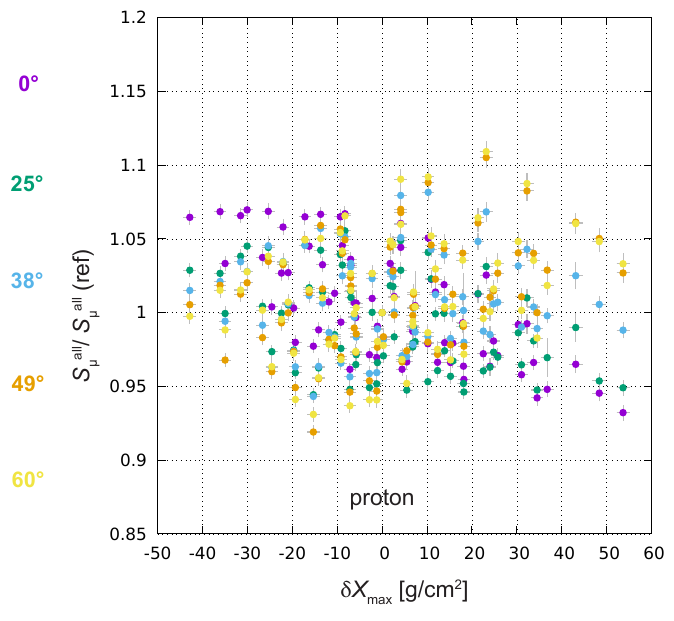}
\vskip-3mm\caption{Relative change in the number of all muons at ground and shift of $X_\mathrm{max}$ for all considered modifications, color-coded for different zenith angles.}
\label{fig1}
\end{figure}

\begin{figure}
\vskip1mm
\includegraphics[width=\column]{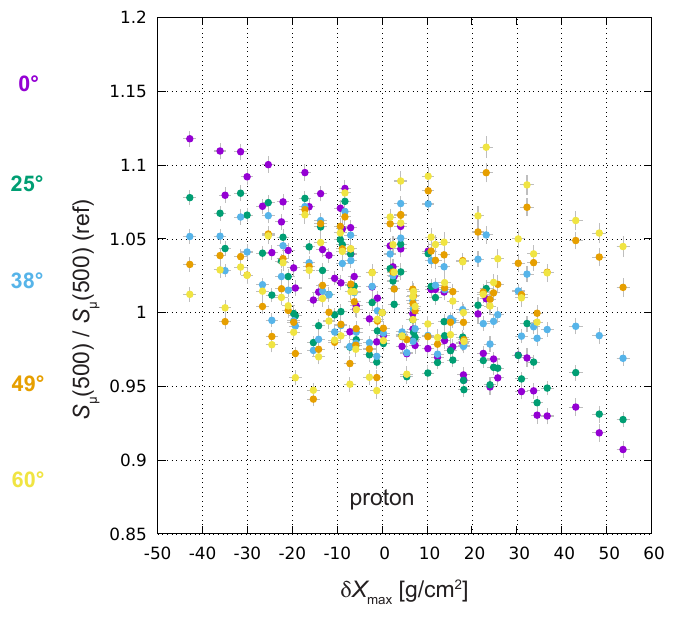}
\vskip-3mm\caption{Relative change in the number of muons at ground 500 meters from the shower core and shift of $X_\mathrm{max}$ for all considered modifications, color-coded for different zenith angles.}
\label{fig2}
\end{figure}

\begin{figure}
\vskip1mm
\includegraphics[width=\column]{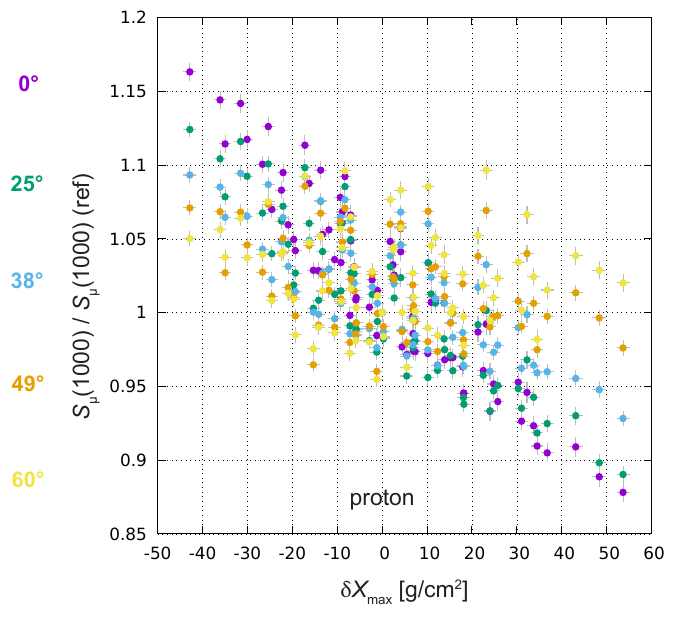}
\vskip-3mm\caption{Relative change in the number of muons at ground 1000 meters from the shower core and shift of $X_\mathrm{max}$ for all considered modifications, color-coded for different zenith angles.}
\label{fig3}
\end{figure}

\begin{figure}
\vskip1mm
\includegraphics[width=\column]{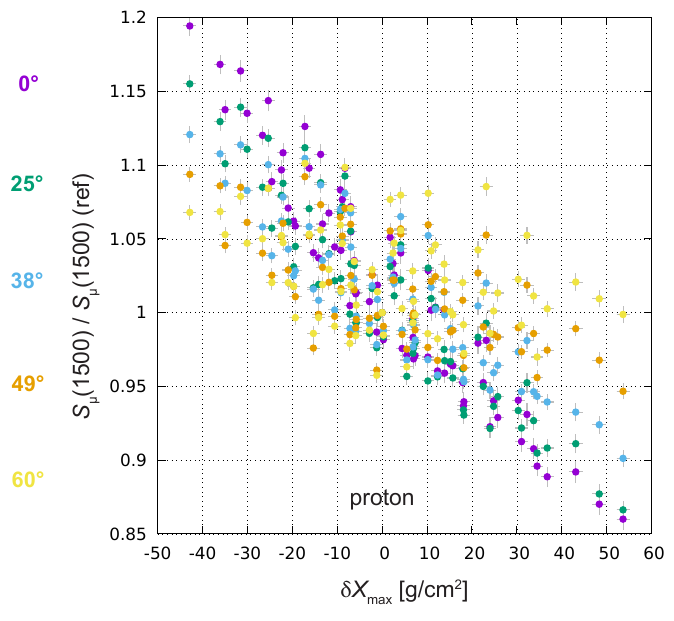}
\vskip-3mm\caption{Relative change in the number of muons at ground 1500 meters from the shower core and shift of $X_\mathrm{max}$ for all considered modifications, color-coded for different zenith angles.}
\label{fig4}
\end{figure}

\section{Introduction}

Ultra-high-energy cosmic (UHECR) rays arrive to the Earth at energies up to the order of $10^{20}$ eV, which, for primary protons interacting with single nucleons in the air, corresponds to over 400 TeV in the center-of-mass system, far surpassing the energies reached at current accelerators. The resulting extensive air showers are simulated using hadronic interaction models tuned to accelerator data -- the most up-to-date models suitable for air shower simulations are Sibyll 2.3d \cite{sibyll}, QGSJETII-04 \cite{qgs} and EPOS-LHC \cite{epos}. The muon puzzle \cite{puzzle} -- the observation of more muons arriving at ground in UHECR showers than what is predicted by the simulations -- has been a long-standing issue in the field. Recently, an analysis of the data of the Pierre Auger Observatory \cite{vicha} has indicated that the simulation underestimate not only muon number at ground but also the depth of the maximum of the development of the air showers in the atmosphere $X_\mathrm{max}$.

Since the properties of hadronic interactions cannot be calculated from first principles, all hadronic interaction models are based on a principally phenomenological description of the underlying physics and thus the extrapolations of their predictions beyond the accelerator energies are uncertain. The three main models are based on similar concepts -- such as Pomeron exchange, Gribbov-Regge theory, minijets, string fragmentation etc. -- but are deeply different from each other. Yet it is clear that three specific models, with particular sets of tuned parameters, do not even remotely cover the entire landscape of possibilities for the actual properties of the hadronic interactions -- if we find one of the models fitting particular data better, it is still difficult to make any conclusions from that fact about any properties of the interactions.

\section{Modified hardonic interactions in 3D}

It has been realized already over a decade ago that a very useful way to gain insight into the hadronic interactions at ultra-high energies is to carry out simulations using the standard interaction models but with ad hoc modifications of some major properties of the interaction. This has been extensively explored in the work \cite{ulrich}, where such modifications were implemented for interaction cross-section, elasticity, multiplicity and charge ratio (the ratio of charged to neutral pions) within the CONEX simulation framework \cite{conex}. However the 1-dimensional nature of the CONEX package severely limits the potential for comparison with experimental data for ground observables. Due to the very low flux of UHECR, any ground observatory hoping to detect any significant number of them needs to cover a very large area -- such observatory is then naturally sparse and the signal for most showers, the signal is typically sampled at least hundreds of meters away from the shower core.

\begin{figure*}
\vskip1mm
\includegraphics[width=\textwidth]{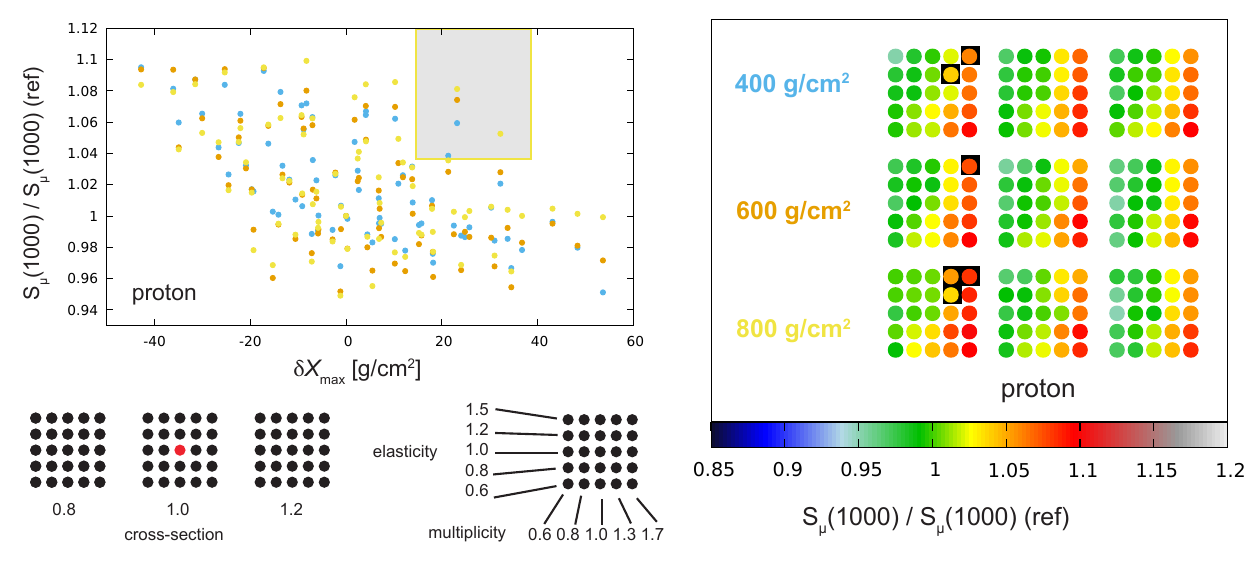}
\vskip-3mm\caption{Relative change of muon signal at 1000 m for three different distances between the ground and the shower maximum D$X$ for primary protons. Left top: correlation with mean $X_{\rm max}$ shift for all considered modifications. The shaded area corresponds to the results of the Pierre Auger Observatory at $\theta=55$ \cite{vicha}. The modifications that fall within this area are highlighted in the right panel -- the legend to the visualisation of the 75 different combination of modifications is shown on the bottom left part of the plot, this pattern is repeated in the right panel three times for the three different D$X$ values.}
\label{fig5}
\end{figure*}

\begin{figure*}
\vskip1mm
\includegraphics[width=\textwidth]{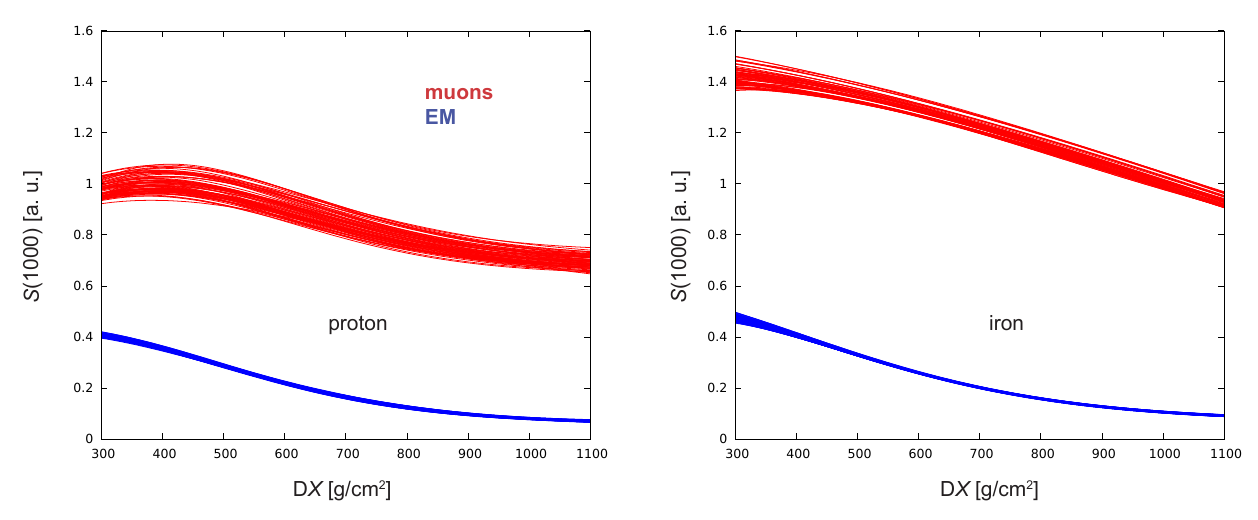}
\vskip-3mm\caption{Dependence of the energy density of EM particles and number density of muons in 1000 meters from the shower core (expressed as ground signal in arbitrary units) on the distance between the ground and the shower maximum D$X$ for primary protons and iron nuclei.}
\label{fig6}
\end{figure*}

\begin{figure*}
\vskip1mm
\includegraphics[width=\textwidth]{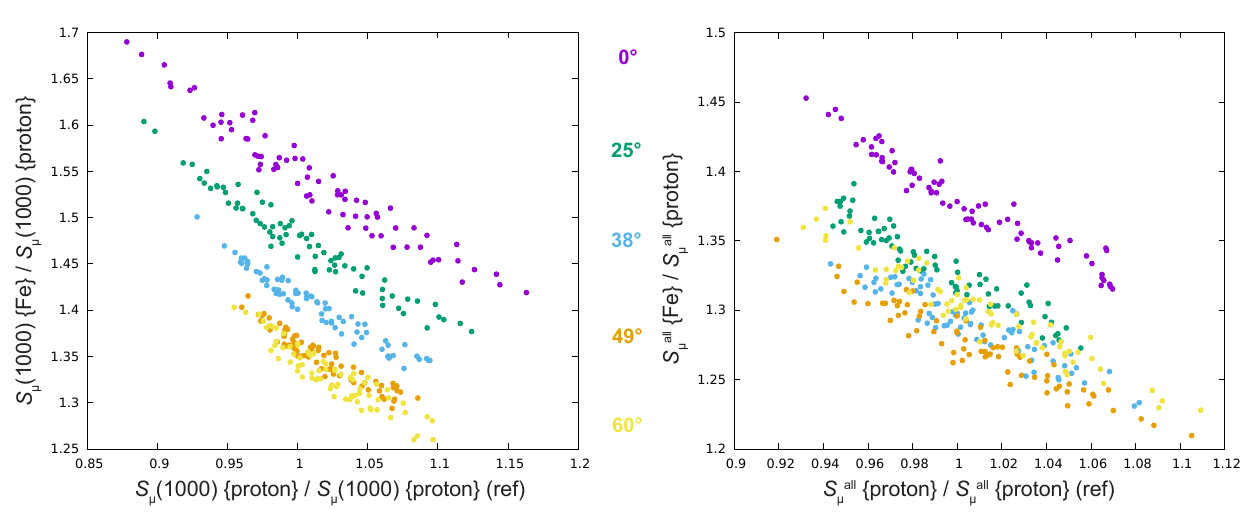}
\vskip-3mm\caption{Ratio of muon number between iron and proton showers for all considered modifications and different zenith angles (color coded) as a function of the change in the mueon number for protons -- for muons at 1000 meters from the shower core in the left panel and for all muons in the right panel.}
\label{fig7}
\end{figure*}

We have thus implemented the same modifications of interaction cross-section, elasticity and multiplicity in the full 3-dimensional simulation framework CORSIKA \cite{corsika}. The technical details of the implementation are described in \cite{mochi1,mochi2,mochi3} - here we only summarize that we are modifying the properties of every interaction simulated by Sibyll 2.3d above a given threshold ($10^{16}$ eV, $10^{15}$ eV, $10^{14}$ eV for cross-section, multiplicity and elasticity respectively) and the magnitude of the modifications grows logarithmically with energy from 1 (no modifications) at the threshold to a given value $f_{19}$ at $10^{19}$ eV, where the values of  $f_{19}$ are (0.8,1,1.2) for cross-section, (0.6,0.8,1,1.3,1.7) for multiplicity and (0.6,0.8,1,1.2,1.5) for elasticity, thus creating 75 different combinations of modified simulations, which are then carried out for proton and iron primaries at $10^{18.7}$ eV for 5 zenith angles. At 1000 showers per a specific combination of modifications, primary particle and zenith angle, this "MOCHI library" contains 750 thousand showers.

It is important to note that the modifications are implemented for a nucleon-air interactions -- in the case of nuclear projectiles (primary iron and its secondary products), we take advantage of the fact that Sibyll treats the individual nucelon-air interactions individually and apply the modifications to them. This also means that the primary energy is shared between the projectile nucleons and thus the thresholds are effectively $A$ times higher for nuclear projectiles.

The importance of the 3-dimensional approach is easily seen from Figs.~\ref{fig1}--\ref{fig4}, which show the changes in the number of muons at ground (taken here at 1400 meters above the sea level, the altitude of the Pierre Auger Observatory) and the depth of the shower maximum $X_\mathrm{max}$ for all of the considered modifications, for proton primaries. The results are shown for all muons in Fig.~\ref{fig1} and then for muons at increasing distances from the shower core for the remaining three figures. The difference between the change in the number of muons at a significant distance from the shower core (which can only be obtained from a 3-dimensional simulation) and in the total number of muons (accessible through 1-dimensional simulations) is large particularly for low zenith angles.

Since the Pierre Auger Observatory reports the ground signal at 1000 meters from the shower core, the 3-dimensional simulation is crucial for the comparison with its results. To compare our simulations with the Auger analysis \cite{vicha}, we must first correct the change in the number of muons at ground for the change induced by the shift of $X_\mathrm{max}$; this correction is conveniently accounted for by using the difference D$X$ between $X_\mathrm{max}$ and the slant depth of the ground (see \cite{mochi1} for details). Fig.~\ref{fig5} shows the results for three different relevant values of D$X$ compared with the range of changes of the number of muons and $X_\mathrm{max}$ indicated by Auger. Interestingly, only specific combinations of extreme values of $f_{19}$ for all three modifications (decreased cross-section and increased elasticity and multiplicity) reach this area.

\section{Further insights}

The range of modifications considered was chosen so that the available  constraints from accelerator data are not violated. This does not necessarily mean that all combinations of the modifications represent an equally realistic option for the description of the real hadronic interactions, since further cosmic ray data should be taken into account. However the entire MOCHI library offers an interesting resource, because it provides a set of interaction models with a much broader range of differences than there is between just the three major interaction models. It can thus be used to better estimate systematic uncertainties do to the unknown hadronic physics. Moreover, it can be also used to study the range of validity of various assumptions about hadronic interactions, such as the correlations between different variables and any features believed to hold independently of the details of the interaction. In this section, we present two examples of such applications.

At UHECR energies, the extensive air showers develop over so many generations that the majority of the primary energy is eventually transferred to the electromagnetic (EM) component of the shower (electrons, positrons and photons). Interestingly, the EM component of the showers is largely universal - the development of the EM component depends on the details of hadronic interaction chiefly through the changes in the depth of its maximum $X_\mathrm{max}$ and thus the ground energy density of EM particles depends mostly only on the distance D$X$ between the ground and the maximum, at least for those EM particles that come from high-energy neutral pion decays. This idea has been formalized in the form of the "shower universality" \cite{ave} and has been proven to hold in simulations to a high precision if additional details of the EM component are taken into account  -- but only for a limited number of hadronic interaction models.

In Fig.~\ref{fig6}, we show the dependence of EM energy density and muon number on D$X$ for all of the 75 combinations of modifications. It is clear that the universality of the EM component holds very well, even  for extreme combinations of modifications, and that the EM component develops similarly for proton and iron. The muon number, on the other hand, is strongly sensitive to both the primary mass and the modification hadronic interactions -- its weaker dependence on the modifications for iron is due to the treatment of nuclear-air interactions as a superposition of nucelus-air interactions at $A$-times smaller energy, leading to smaller modifications.

This "superposition model" approach has another important consequence, illustrated in Fig.~\ref{fig7}. The "muon puzzle" in UHECR data requires that additional muons are added in simulations. A given combination of modifications will, due to the lower effect on nuclei, always add more muons to proton-induced showers than to iron-induced ones. However since heavier nuclei already produced more muons, any modifications that increase the number of muons will naturally decrease the ratio between the number of muons predicted for iron and proton primaries. Such a trend is observed already for the total number of muons, but the 3D simulations allow us to quantify it for any distance from the core. 

This observation is significant for any UHECR observatory that aims to distinguish the primary particles based on the muon content of the showers, because the sensitivity is typically assessed using simulations with the current hadronic interaction models. If these models do indeed  underestimate the number of muons in the showers, they may overestimate the separation power of the experiment. In our implementation, this effect is a direct consequence of our choice to treat the modifications for nuclei as a superposition of modifications for individual nucleons. While in the real interactions, this approximation may not be exactly valid, the superposition model of hadronic interactions has been largely successful in explaining many phenomena at ultra-high energies and it is expected to hold at least to some extent, unless radically new physics is present (in which case the entire approach presented here is not directly applicable).

\vskip3mm \textit{Acknowledgement.} This work was co-funded by the EU and supported by the Czech Ministry of Education, Youth and Sports through the project CZ.02.01.01/00/22\_008/0004632, GACR (Project No. 21-02226M) and CAS (LQ100102401).

\end{document}